\documentclass[11pt,twoside]{article}
\usepackage{asp2004}
\usepackage{psfig}
\usepackage{epsf}
\usepackage{graphics}
\usepackage{lscape}
\def\edcomment#1{\iffalse\marginpar{\raggedright\sl#1\/}\else\relax\fi}
\reversemarginpar

\begin{document}
\title{Redshift Dependence of QSO Spectral Variability}
\author{F. Vagnetti}
\affil{Universit\`a di Roma Tor Vergata, Dipartimento di Fisica, via della Ricerca Scientifica 1,
I-00133 Roma, Italy}

\author{D. Trevese}
\affil{Universit\`a di Roma La Sapienza, Dipartimento di Fisica, p.le Aldo Moro 2, I-00185 Roma, Italy}

\begin{abstract}
We present a combined analysis of the optical spectral
variability for
two samples of QSOs, 42 objects at $z<0.4$ monitored at the Wise
Observatory \citep{giv}, plus 59 objects up to $z\sim 3$ in the
field of the Magellanic Clouds, detected and/or monitored within the MACHO
Project database \citep{geh}. Our analysis shows some increase of
the observed spectral variability as a function of redshift, with a
large scatter. These data are compared with a model based on the addition
of flares of different temperatures to a stationary quasar SED,
taking into account also the intrinsic scatter of the SEDs.
\end{abstract}
\thispagestyle{plain}

\section*{Introduction}
Variability of the spectral energy distribution (SED) of Active
Galactic Nuclei (AGNs) is a powerful tool to investigate the role of
the main emission processes, and the origin of their variations. Many
variability mechanisms have been proposed in the past, including
supernovae explosions \citep{are}, instabilities in the
accretion disk \citep{kaw}, and gravitational lensing due
to intervening matter \citep{haw}. Since it is likely that
different physical phenomena are causing variability in different
bands and timescales, multifrequency analyses will be ultimately
needed to obtain a complete description. However, we have shown in a
previous paper \citep{tv2} that even a two optical band
analysis, once performed on a statistical sample, provides valuable
constraints on the origin of variability. In the optical band, AGNs
most commonly become bluer, i.e. their spectrum becomes harder, when
brighter. This has been shown to occur for some individual AGNs \citep{cut,ede,kin,pal},
and for a few complete samples: PG quasars at
$z<0.4$ monitored at the Wise Observatory \citep[hereafter referred as WO quasars]{giv},
faint quasars in the SA 57 with
two-epoch information \citep{tkb}, and quasars
from the MACHO Project database \citep[hereafter referred as MP quasars]{geh},
which are here studied together with the WO quasars \citep[see also][]{vt3}.
Our analysis \citep{tv2} of the two-color WO light curves confirms the
hardening in the bright phase of individual quasars. This in turn accounts
for two other effects: (i) average variability of quasar samples is greater at
higher rest-frame frequencies (Di Clemente et al. 1996), a results confirmed
by recent results of \citet{van}, \citet{dev03}, \citet{dev04},
and (ii) -- at a given observed frequency -- variability is correspondingly
greater at higher redshifts \citep{gtv}.
To quantify the amount of spectral variability, we \citep{tv2}
introduced the spectral variability parameter
$\beta\equiv\Delta\alpha/\Delta\log f_{\nu}$
($ f_{\nu}$ being the
specific flux and $\alpha\equiv\partial\log f_{\nu}/\partial\log\nu$
the spectral slope) and the use of the $\alpha$-$\beta$ plane to
derive constraints on the variability mechanisms. In particular, it
was shown that the variation of the spectral slope implied by changes
of the accretion rate is not able to explain $\beta$ values deduced
from the observations, while ``hot spots'' on the accretion disk,
possibly caused by instability phenomena \citep{kaw}, can
easily account for the observed spectral variability.

\section*{Samples} 
The present analysis is based on the sets of light curves for the Wise
Observatory (WO) sample \citep{giv} and for the MACHO Project
(MP) sample \citep{geh}, both available in electronic form.
The WO sample consists of 42 PG quasars selected to be  nearby, i.e. $z <
0.4$, and bright, i.e. $B < 16$ mag, monitored in the Johnson-Cousins B
and R bands with the 1 m Wise Observatory telescope for a total duration
of 7 years and a median observing interval of 39 days.
The MP sample includes 47 quasars discovered behind the
Magellanic Clouds on the basis of their variability, and 12 additional
quasars previously known in the same area. All the 59 MP quasars were
monitored for 7.5 years with Mount Stromlo Observatory's 1.27 m
telescope, with average sampling times of 2-10 days. Magnitudes in
the special MACHO red and blue passbands were then transformed and made
available as standard V and R magnitudes.
The MP quasars cover the redshift interval $0.2<z<2.8$ and have apparent
mean magnitudes $16.6\leq{\bar V}\leq20.1$.
The combined analysis of the two samples allows the investigation
of the spectral variability as a function of $z$.

\section*{Single Band Variability} 
Analysis of the dependence of variability on redshift and luminosity
on single magnitude-limited samples is affected by the strong $L$-$z$
correlation intrinsic to the samples themselves, therefore it is
preferable to combine more samples to cover a larger portion of the
$L$-$z$ plane \citep[see e.g.][]{gtv}. Thus, we
combine the MP sample, with data in V and R for 59 quasars, and the WO
sample, which has variability information in B and R for 42 PG quasars
at $z\la 0.4$. Following \citet{gtv}, we
measure variability in a fixed bin of the rest-frame Structure Function,
75-150 days. For the R band, which is common to the two samples, we find
no correlation of variability with redshift for the MP sample alone, but
we find an increase of variability with redshift for the combined
samples, with correlation coefficient $r=0.19$ and probability
$P(>r)=0.05$, and a more significant result, $r=0.44, P(>r)=3\cdot
10^{-4}$, if we consider only objects of the two samples in a limited
range of absolute magnitudes ($-26<M_R<-22$). We find also that
variability increases with $M_R$, with $r=0.25, P(>r)=0.01$ for the
combined samples, and with $r=0.30, P(>r)=0.02$ for the MP sample alone.
These results confirm the previous trends found, e.g., by \citet{gtv}
and by \citet{cri}.
Variability increases also as a function of rest-frame frequency \citep{dic};
results for the MP and WO samples have been shown by \citet{vt3}.

\begin{figure}[!ht]
\plottwo{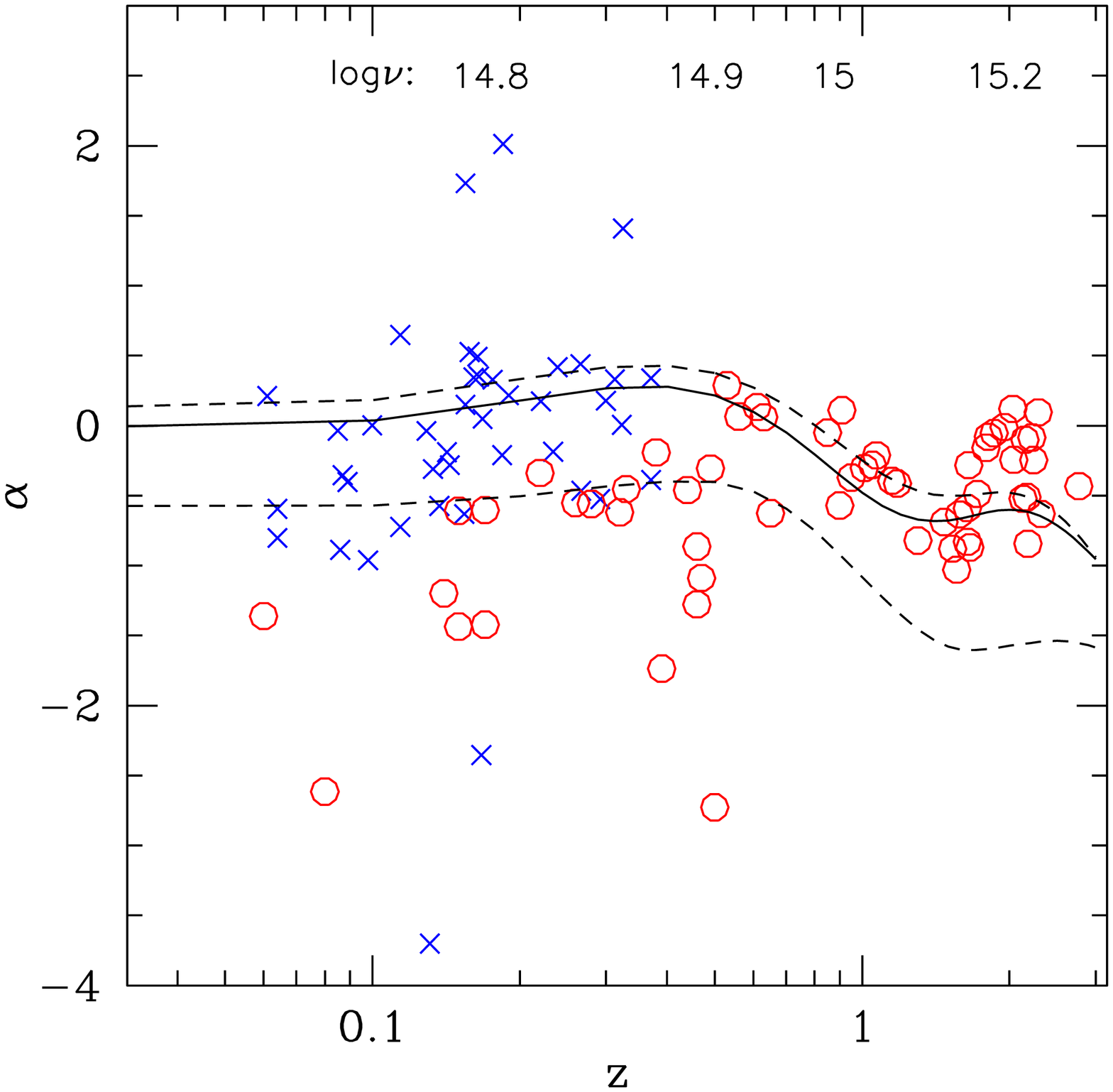}{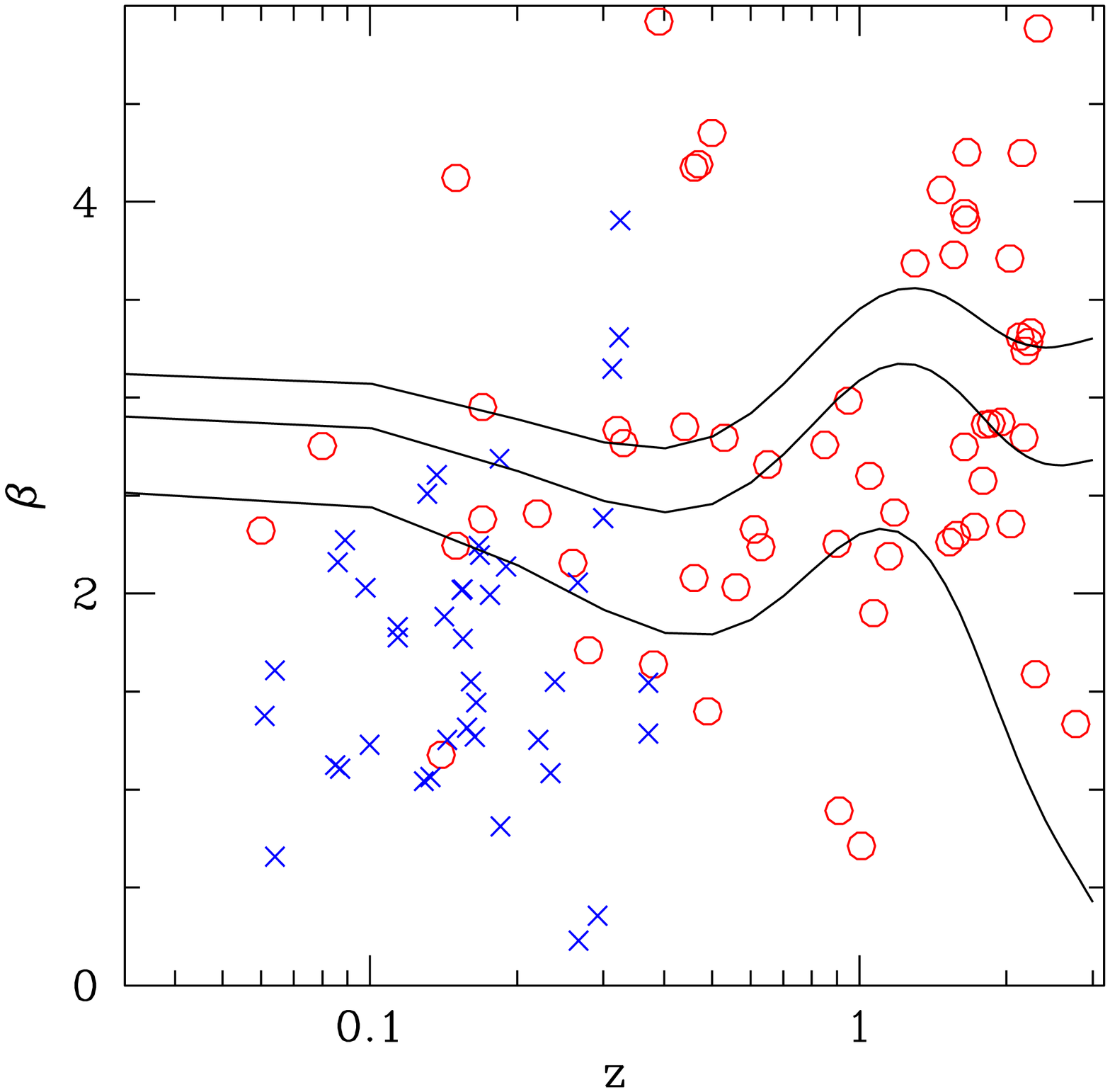}
\caption{{\it Left:} spectral slope $\alpha$ vs $\log z$ for the two quasar
samples; crosses: WO sample; circles: MP sample. The continuous curve
represents the prediction for the average quasar SED determined by \citet{elv},
assumed independent of $z$ in shape and plotted as a function of the redshift corresponding to $\nu_{rest}$ (which is labeled 
in the upper part of the diagram) for the observed V band.
Dashed curves correspond to the
90\% percentile SEDs published by the same authors. 
{\it Right:} The spectral variability parameter $\beta$ vs $\log z$ for the two
samples. The curves correspond to model predictions for the average SED
by \citet{elv}, perturbed with blackbody flares of temperatures
$T=2\cdot 10^5$ K (upper curve), $T=6\cdot 10^4$ K (middle curve),
$T=3\cdot 10^4$ K (lower curve).}
\end{figure}

\section*{Spectral Variability}
As the MP sample includes objects in a large range of redshift, up to
$z\la 3$, the spectral slope $\alpha$ measured at a given observed
frequency is in fact the slope of the SED at different emission
frequencies, increasing with $z$. Therefore we plot, in Figure 1 ({\it left})
$\alpha$ vs $z$ for the two samples, compared with a prediction for an
average SED, as determined by \citet{elv}. To take into account the
large scatter of the quasar SEDs, we also show the predictions for the
90\% percentiles reported by \citet{elv}. The comparison is
consistent with the assumption of an average SED approximately
independent of $z$ in shape,
although some larger scatter in the spectral slopes
appears in the low-$z$ data.
To perform a similar comparison for the spectral variability parameter
$\beta$, as a function of $z$, we require a model for the
variability of the SED; we adopt, as in \citet{tv2}, the
addition of a blackbody flare to the average SED by \citet{elv}.
In Figure 1 ({\it right}), showing $\beta$ vs
$z$ for the two samples, some increase of $\beta$ with
$z$ is apparent, but there is a large scatter. 
The model predictions, for flares of different temperatures added to the
average quasar SED, account for at least part of the scatter.

\section*{Conclusions} 
The study of spectral variability of quasars is a powerful tool to
test the validity of different emission and variability models against
time-dependent data of statistical quasar samples. For such purpose,
it is necessary to obtain variability information in at least 2
bands. The samples of quasars by \citet{giv}, with light curves
for 42 quasars in B and R, and that by \citet{geh}, with information
in V and R for 59 quasars, are both appropriate for this purpose.
Moreover, the latter sample allows to extend investigation of the spectral
variability to high $z$.
Since the MP sample is selected on the basis of variability, our analysis
could be biased by possible systematic differences in the variability
properties of the two samples. However, since the variability threshold
adopted for detection in the MP sample is low (0.05 mag), the incompleteness
respect, e.g., to color-selected samples becomes negligible \citep[see][]{t89}.
Investigation of spectral variability as a function of redshift
corresponds to the study of the perturbations of the SED at different
rest-frame frequencies. Large samples with multi-band variability
information are ultimately needed to characterize the dependence of the
spectral variability parameter $\beta$ on both $\nu_{rest}$ and $z$.
Progress in this direction will be achieved by a new
analysis of the type presented by \citet{van},
if repeated observations will provide a better time sampling.

\end{document}